\documentclass[aps,floatfix,prd,preprint]{revtex4}
\usepackage{graphicx} 
\usepackage{bm}
\usepackage{amsmath}
\usepackage{float}
\usepackage{longtable}
\usepackage{braket}
\usepackage{comment}

\def\be{\begin{equation}}
\def\ee{\end{equation}}
\def\A{{\bf A}}
\def\B{{\bf B}}
\def\x{{\bf x}}
\def\r{{\bf r}}
\def\y{{\bf y}}

\def\D{{\bf D}}

\newcommand{\bnabla} {{\mbox{\boldmath $\nabla$}}}

\begin{document}

\title{A Constituent Model of Light Hybrid Meson Decays}
\author{Christian Farina}
\affiliation{Department of Physics and Astronomy, University of Pittsburgh, Pittsburgh, PA 15260, USA.}
\author{Eric S.\,Swanson}\email[]{swansone@pitt.edu}
\affiliation{Department of Physics and Astronomy, University of Pittsburgh, Pittsburgh, PA 15260, USA.}

\begin{abstract}
A model of light hybrid mesons and their strong decays is developed. The model employs a gluonic quasiparticle to describe low energy gluodynamics and uses the QCD Hamiltonian in Coulomb gauge to guide the construction of states and decay amplitudes. We compute the  partial widths of the twelve low lying isovector and vector hybrids. Implications of these results on hybrid searches are also made, with the chief conclusions being that direct observation of the vector states will be difficult, that a hybrid $\pi(1800)$ has distinctive decay characteristics, a narrow $\eta(1900)$ hybrid should exist,  an $\eta_1(1750)$ should be sought, and that the exotic nature of $J^{PC}= 2^{-+}$ hybrid mesons should be discernible with sufficient data. We argue that the isovector $\pi_2$ hybrid has been discovered, giving a total of four possible hybrid mesons, $\pi_1(1600)$, $\eta_1(1855)$, $\pi(1800)$, and $\pi_2(2360)$, which appear to be filling out the lowlying hybrid supermultiplet in the expected fashion.
\end{abstract}

\maketitle

\section{Introduction}

The search for gluonic excitations in hadronic matter has been an ongoing process for many decades. The advent of GlueX, the possibility of light hybrid production in charmonium decay at facilities such BEPC, and future prospects with the  PANDA experiment make understanding possible decay modes of hybrids especially useful and timely. Of course, modeling such decays requires an appropriate picture of the nature of a hybrid meson. While earlier work has relied on intuitive notions of nonperturbative gluons as strings or some type of quantum particle (for a review see Ref. \cite{Meyer:2015eta}), recent lattice field computations convincingly demonstrate that the low lying charm hybrid spectrum can be understood in terms of a gluonic quasiparticles with axial quantum numbers\cite{Dudek:2011bn}. This observation meshes well with an earlier hybrid model that uses the Hamiltonian of QCD in Coulomb gauge as a starting point\cite{Swanson:1998kx,Swanson:1997wy,Guo:2008yz}.

A central element of the latter approach is the use of a gluonic vacuum ansatz that gives rise to a gap equation with a quasiparticle solution that can be thought of as a gluon with  dynamical mass\cite{Szczepaniak:2001rg}. This ``mass" then makes the construction of a Fock space sensible. An additional feature of the formalism is that the interactions of the constituents are completely specified in terms of the QCD Hamiltonian. A recent computation of flavor mixing in light hybrids shows reasonable agreement with lattice field results, indicating that the approach is not without merit\cite{Swanson:2023zlm}.

The goal of this investigation is to compute the decays of light isovector and isoscalar hybrid mesons to pairs of mesons. Thus, we construct hybrid mesons with constituent quarks and gluons and permit their strong decay with the leading operator in the QCD Hamiltonian. We find total decay widths ranging from tens of MeV to hundreds of MeV. Our  result for the $J^{PC} = 1^{-+}$ isovector agree well with experimental values for the $\pi_1(1600)$ hybrid candidate and with lattice computations for the decay properties of this state. Prospects for hybrid discovery and diagnostics for their detection are discussed in Section \ref{sect:conc}, where we conclude that the vector hybrid is too broad for direct detection, the $\eta_1(1855)$ is the likely $s\bar s g$ partner to the $\pi_1(1600)$,  that an $\eta_1(1750)$ light partner state is expected, and find diagnostic decays for a possible $\pi(1800)$ hybrid. We also suggest that the $\pi_2$ hybrid state has already been sighted.

\section{Hybrid Modelling with Constituent Gluons}

The Hamiltonian of QCD in Coulomb gauge is the starting point for our model of hybrids. This gauge choice is useful for model building because Gauss's law is resolved, all degrees of freedom are physical, and an explicit interaction potential that operates between quarks and gluons emerges. This instantaneous interaction is obtained in the same way as the Coulomb interaction in QED and can be written as

\begin{equation}
V_C = \frac{1}{2}\int d^3x\, d^3y\, {\cal J}^{-1/2} \rho^A(\x)
 {\cal J}^{1/2} \hat K_{AB}(\x,\y;\A) {\cal J}^{1/2} \rho^B(\y) {\cal J}^{-1/2},
\label{eq:hc}
\end{equation}
where the Faddeev-Popov determinant is written as ${\cal J} \equiv {\rm det}(\bnabla\cdot \D)$ and $\D$ is the adjoint covariant derivative, 
 $\D^{AB} \equiv \delta^{AB} \bnabla  - g f^{ABC}\A^C$.
The color charge density included quark and gluonic components and is given by

\begin{equation}
\rho^A({\bf x}) =
 f^{ABC} {\bf A}^B({\bf x}) \cdot {\bf \Pi}^C({\bf x}) + \psi^{\dag}(\x)T^A\psi(\x).
\label{eq:rho}
\end{equation}
The kernel of the Coulomb interaction can be formally written as\cite{Christ:1980ku}

\begin{equation}
\hat K^{AB}({\bf x},{\bf y};\A) \equiv \langle{\bf x},A|
 \frac{ g }{ \bnabla\cdot {\bf D}}(-\bnabla^2)
 \frac{ g }{ \bnabla\cdot {\bf D}}|{\bf y},B\rangle.
\label{eq:K}
\end{equation}
Finally, $\A$ is the vector potential and ${\bm \Pi}$ is the conjugate momentum given by the negative of the transverse chromoelectric field.
The Coulomb interaction, along with the quark and gluon kinetic energies, gluon self-interactions, and the quark-transverse gluon interaction, $-g \int d^3x \psi^\dagger \bm{\alpha}\cdot \A \psi$, comprise a full field-theoretic version of QCD, with its accompanying nonperturbative features. 

A quasigluon that is consistent with the constraints of QCD can be developed with a mean field model of the gluonic vacuum. The resulting Schwinger-Dyson equations yield an estimate for the vacuum expectation of the kernel and for the gluon dispersion relationship. This, in turn, can be used to evaluate the expectation value of $\hat K^{AB}$, which mimics linear confinement interactions of quark models and lattice field theory\cite{Szczepaniak:2001rg}. Because of this, we choose to simply model the vacuum expectation of the Coulomb kernel as:

\be
\langle \hat K^{AB}(\r,\A)\rangle \to
\delta^{AB}\left( -\frac{3}{4} \mathcal{C} + \frac{a_S}{r} - \frac{3}{4} \sigma r\right).
\label{eq:V}
\ee
Of course this reproduces the successes of the Cornell potential in nonrelativistic quark models and agrees with lattice computations of the Wilson loop potential.

As mentioned, the vacuum model also gives rise to a quasigluon with a dispersion relation that can be approximated as\cite{Szczepaniak:2001rg}

\be
\omega^2 = k^2 + m_g^2\, \exp(-k/b_g).
\label{eq:omega}
\ee
The dynamical gluon mass is $m_g \approx 600$ MeV and the parameter $b_g \approx 6000$ MeV. We stress that the gluon remains transverse and properties, such as Yang's theorem, remain in place. Other vacuum Ans\"{a}tze are possible, for example a Gaussian wavefunctional (equivalent to the mean field approximation described) can be combined with the Faddeev-Popov operator, which gives rise to a dispersion relation that is well-described by the Gribov form, $\omega^2 = k^2 + m_g^4/k^2$\cite{Feuchter:2004mk}. Both forms will be considered in the following.

With these elements in place, one obtains a comprehensive microscopic model of hadrons and their interactions that 
can be thought of as a minimal extension of the constituent quark model with the addition of constituent gluon degrees of freedom and  many-body interactions that are contained in the expansion of the Coulomb 
operator of Eq. \ref{eq:K}.

\subsection{Light Hybrids}

Construction of a hybrid meson is conveniently done by 
coupling the gluon spin $j_g$ to the gluon angular momentum, $\ell_g$. Converting to the gluon helicity basis and assuming  that  $\ell_g = j_g$ produces a factor of

\be
\chi^{(-)}_{\lambda,\mu} \equiv \langle 1 \lambda \ell_g 0| \ell_g \mu\rangle =
\begin{cases} 0, \ell_g = 0 \\ \frac{\lambda}{\sqrt{2}} \delta_{\lambda,\mu}, \ell_g \geq 1 \end{cases}.
\ee
This represents a transverse electric (TE) gluon in our model and forms the explicit realization of the axial constituent gluon. Alternatively, one may set $\ell_g = j_g\pm 1$ and obtain a transverse magnetic (TM) gluon with a Clebsch factor given by $\chi^{(+)}_{\lambda,\mu} = \delta_{\lambda,\mu}/\sqrt{2}$.  Combining with quark spins yields the final expression for a hybrid creation operator

\begin{align}
&|JM [LS \ell j_g \xi]\rangle = \frac{1}{2} T_{ij}^A\int \frac{d^3q}{(2\pi)^3}\, \frac{d^3k}{(2\pi)^3} \,
\Psi_{j_g;\ell m_\ell}({\bf k}, {\bf q})\, \sqrt{\frac{2 j_g +1}{4\pi}} \, D_{m_g\mu}^{j_g*}(\hat k) \, \chi^{(\xi)}_{\mu,\lambda} \nonumber \\
& \times
\langle \frac{1}{2} m \frac{1}{2} \bar{m} | S M_{S} \rangle \,
\langle \ell m_\ell, j_g m_g| L M_L\rangle \,
\langle S M_S, L M_{L} | J M \rangle \,
b_{{\bf q}-\frac{{\bf k}}{2},i,m}^\dagger \,
  d_{-{\bf q}-\frac{{\bf k}}{2},j, \bar{m}}^\dagger \,
a^\dagger_{{\bf k}, A, \lambda} |0\rangle.
\label{eq:PSI}
\end{align}
As discussed, the quark and gluon operators create quasiparticles and $|0\rangle$ refers to the correlated vacuum. By construction, the hybrid state is an eigenstate of parity and charge conjugation with eigenvalues given by

\be
P =\xi (-1)^{\ell + j_g+1}  \ \ \textrm{and} \  \ C= (-1)^{\ell+S+1}.
\ee
Lowest lying hybrid states have $\xi=-1$ and $j_g=1$, which we assume from now on.

\subsection{Model Parameters}
\label{sect:model}

Hybrid decay amplitudes require wavefunctions for the initial hybrid and final ordinary meson states. These are obtained by solving the  associated bound state equations, which in turn depend on the interaction parameters of Eq. \ref{eq:V}, the quark masses, and parameters describing  spin-dependent interactions, if desired. An example interaction is obtained from Ref.~\cite{Swanson:2023zlm}:

\begin{align}
V_{SD} &= 2C_F \frac{\alpha_H}{3 r m_q m_{\bar q}}\,  S_q \cdot S_{\bar q} \, b_0^2\exp(-b_0 r) \, \theta(r_0)\nonumber \\
       &+ \frac{1}{2}\left(C_F \frac{\alpha_H}{r^3}  + (2\epsilon-1)\frac{b}{r}\right) \cdot \left(\frac{S_q\cdot L}{m_q^2} + \frac{S_{\bar q}\cdot L}{m_{\bar q}^2}\right)\, \theta(r_0) + \left(C_F\frac{\alpha_H}{r^3} + \epsilon\frac{b}{r}\right) \frac{S\cdot L}{m_q m_{\bar q}}\, \theta(r_0) + \nonumber \\
       &+ \frac{4\alpha_H}{3 m_q m_{\bar q} r^3} \left( S_q\cdot r \, S{\bar q}\cdot r - S_q \cdot S_{\bar q}\right) \, \theta(r_0).
\end{align}
The notation $\theta(r_0)$ denotes infrared regulation by saturating the expressions at $r_0$ whenever $r<r_0$. The parameter $\epsilon$ summarizes mixing in the Lorentz structure of the long range spin-dependent interaction and is discussed further in Ref. \cite{Swanson:2023zlm}. All parameters are obtained by fitting light meson masses. 

In an effort to gauge model sensitivity, we have performed 
fits under a variety of conditions: for example we have fit the masses of 8 $s\bar{s}$ mesons, 23 isovectors, and 44 isoscalar, isovectors, and $I=1/2$ mesons separately. We also varied parameter constraints to obtain several fits to these spectra. A sample of the results are reported in Table \ref{tab:models2}; additional fits can be found in Ref. \cite{Swanson:2023zlm}.

\begin{table}[h]
\caption{Model Parameters. A: fit to $uds$ mesons, A': alternate fit, B: fit to $s\bar s$ mesons, C: fit to $s\bar s$ with $\epsilon, \mathcal{C} =0$.}
\scalebox{0.8}{
\begin{tabular}{l|cccccccc|cc}
\hline\hline
model & $m\ (m_s)$ (MeV) & $a_S$ & $\sigma$ (GeV$^2$) & $\mathcal{C}$ (MeV) & $\alpha_H$ & $b_0$ (GeV$^{-1}$) & $r_0$ (GeV$^{-1}$) & $\epsilon$ (GeV$^{-1}$) & rel error & avg deviation (MeV)\\
\hline
A & 420 (606) & 1.496 & 0.053 & 0.149 & 1.999 & 0.389 & 9.1 & 0.25 & 6\% & 69 \\
A' & 918 (1046) & 1.18 & 0.072 & -0.734 & 8.81 & 0.221 & 8.86 & 0.25 & 6\% & 69 \\
B & (555) & 0.946 & 0.095 & -0.228 & 0.864 & 0.92 & 4.2 & 0.25 & 1.6\% & 21 \\
C & (595) & 0.955 & 0.087 & 0.0 & 1.12 & 0.293 & 4.4  & 0.0 & 7\% & 119 \\
\hline
\hline\hline
\end{tabular}}
\label{tab:models2}
\end{table}

Once parameters are fixed, the light hybrid spectrum can be computed. The hybrid bound state equation is obtained from the  QCD Hamiltonian by computing the expectation value, $\langle J'M'[L'S'\ell' j' \xi']| H | JM[LS \ell j \xi]\rangle$. We seek spin-independent multiplets and therefore consider the nonrelativistic limit of the currents in Eq. \ref{eq:hc}.  The resulting spectrum can be categorized according to interpolating operators, as indicated in Table \ref{tab:JPC}\cite{JKM}. Here $\B$ is the chromomagnetic field, and $\psi$ and $\chi$ are heavy quark and antiquark fields, respectively. The remaining columns give the corresponding quantum numbers in the present model and the hybrid meson quantum numbers in the specified multiplet.

\begin{table}[h]
\caption{$J^{PC}$ Hybrid Multiplets.}
\begin{tabular}{c|c|cccc|l}
\hline\hline
multiplet & operator & $\xi$ & $j_g$ & $\ell$ & $L$ & $J^{PC}\ S=0\ (S=1)$ \\
\hline
$H_1$ & $\psi^\dagger \B \chi$ & -1 & 1 & 0 & 1 & $1^{--}$, $(0,1,2)^{-+}$ \\
$H_2$ & $\psi^\dagger \bnabla \times \B \chi$  & -1 & 1 & 1 & 1 & $1^{++}$, $(0,1,2)^{+-}$ \\
$H_3$ &$\psi^\dagger \bnabla \cdot \B \chi$ & -1 & 1 & 1 & 0 & $0^{++}$, $(1^{+-})$ \\
$H_4$ & $\psi^\dagger [\bnabla \B]_2 \chi$ & -1 & 1 & 1 & 2 & $2^{++}$, $(1,2,3)^{+-}$ \\
\hline\hline
\end{tabular}
\label{tab:JPC}
\end{table}

Calculations were done with the gluon dispersion relationship of Eq. \ref{eq:omega} with $m_g$ set to 600 MeV. Remarkably, using the infrared enhanced Gribov form of the gluon dispersion relationship made little difference in the masses.

The hybrid spectrum was obtained with an expansion in a Gaussian basis and with a novel scheme that is analogous to the Hartree iterative method\cite{Swanson:2023zlm}. Because we are interested in matrix elements that probe the entire wavefunction, the resulting hybrid wavefunctions were approximated by a product of two Gaussians with widths $\alpha$ and $\beta$:

\be
\Psi_{j_g,\ell}(\vec k,\vec q) \propto D^{j_g*}_{m_g,\mu}(\hat k) k \exp(-k^2/2\beta^2) \cdot Y_{\ell m_\ell}(\hat q) q^\ell \exp(-q^2/2\alpha^2)
\ee
Similarly, ordinary mesons are described with SHO wavefunctions described by a scale, $\beta_m$.

Detailed model validation is not feasible at present. A lattice computation of the light meson spectrum at a pion mass of 391 MeV exists, but it has not been able to distinguish enough hybrid states to determine the spin-averaged spectrum\cite{Dudek:2013yja}. This calculation does, however,  set the scale for the spin-averaged $H_1$ multiplet (taken to be the $S=0$, $1^{--}$ mass) to be  $2190\pm 20$ MeV.

We remark that, although the model broadly agrees with lattice field theory calculations, where available, results with charm quarks indicate that further work on improving the spectrum  is required\cite{Farina:2020slb}. This work is under way.

\subsection{Decay Model}

Once hybrid and ordinary meson wavefunctions and masses are determined the decay computation can proceed. We chose to model hadronic decay with the leading operator in the QCD Hamiltonian, which is simply quark dissociation of the valence quasigluon driven by $-g\int d^3x \psi^\dagger\bm{\alpha}\cdot \A \psi$. This has previously been applied to the decay of heavy hybrids in Ref. \cite{Farina:2020slb} and revives an old model due to Tanimoto\cite{Tanimoto}. Although we recognize it as an additional model assumption, the value of the coupling $g$ will be taken to be fixed by $a_S = g^2/(4\pi)$. Details on the numerical techniques used in evaluating the decay amplitudes can be found in Ref.~\cite{Farina:2020slb}.

We note that the decay  model yields selection rules that are typical of hybrid decay models\cite{Page:1996rj}. Namely, spin zero hybrids do not decay to spin zero mesons and  TE hybrids do not decay to mesons with identical spatial wavefunctions.

The chief remaining issue is identifying and modeling the light canonical mesons appearing in the final state. This task is made complicated by mixing in the isoscalar sector, mixing in kaons, difficulty assigning states, and the possible existence of states with multiquark components. For example, it has been suggested that the $f_0(500)$ and $f_0(980)$ can have substantial glueball components \cite{KLEMPT2021136512}, while other authors  argue for their tetraquark nature\cite{Achasov:2020aun}, or for their interpretation in terms of diquarks\cite{Jaffe:1976ig}.
A popular interpretation is that the $f_0(500)$ is strongly influenced by $\pi\pi$ dynamics\cite{pipi}. We thus assume that the $f_0(980)$ is a ground state scalar and that the $f_0(1370)$ is the first radial excitation, while the $f_0(500)$ is roughly modeled with a single broad Gaussian.

Mixing angles for the $\eta$ and $\eta'$ are specified by the simple ansatz:

\begin{eqnarray}
|\eta\rangle &=&\frac{1}{2}(|u\bar u\rangle + | d \bar d\rangle )-\frac{1}{\sqrt{2}}|s\bar{s}\rangle \nonumber \\
|\eta'\rangle &=&\frac{1}{2}(|u\bar{u}\rangle + | d \bar{d}\rangle )+\frac{1}{\sqrt{2}}|s\bar{s}\rangle.
\end{eqnarray}

Spin one kaons are defined as linear superpositions of spin singlet and triplet states through a mixing angle $\theta_{K}$:

\begin{eqnarray}
|K_{1L}\rangle&=&\cos{\theta_{K}}|1P1\rangle+\sin{\theta_{K}}|3P1\rangle \nonumber \\
|K_{1H}\rangle &=&-\sin{\theta_{K}}|1P1\rangle+\cos{\theta_{K}}|3P1\rangle.
\end{eqnarray}
There is much theoretical debate about the value of the mixing angle \cite{Suzuki:1993yc, Cheng:2011pb, Hatanaka:2008gu}, but no experimental data to definitively constrain it. 
The authors of \cite{Cheng:2011pb} argue that the value of $|\theta_{K}|$ can be obtained by first determining the  mixing angles for $f_{1}(1285)/f_{1}(1450)$ and    $h_{1}(1170)/h_{1}(1380)$ from  mass relations. These angles depend on the masses of the $K_{1L}$ and $K_{1H}$, which in turn depend on the mixing angle $\theta_{K}$. This procedure gives a value of $\theta_{K}=35^\circ$. Alternatively, Suzuki determined the mixing angle from  partial decay rates and, independently, from  masses, finding that the angle could be either $\theta_{K}=33^\circ$ or $\theta_{K}=57^\circ$, but that $\theta_{K}=33^\circ$ is favored by the observed production dominance of $K_{1}(1400)$ in  $\tau$ decays\cite{Suzuki:1993yc}. This is the value we employ in the following.

\subsection{Comparison to Lattice Computations}

Ideally the structure and decay models we use could be calibrated against experiment. Unfortunately, very little experimental information is known about hybrids (the present situation is discussed more fully below). There is, however, a lattice QCD computation of the decay properties of the exotic $J^{PC}=1^{-+}$ hybrid to which we can make comparisons\cite{Woss:2020ayi}.

This computation was performed at the ``SU(3)" point, where the three light quark masses are set to the strange quark mass, typically determined by matching to the $\Omega^-$ baryon. Finite volume spectra are then fit to model scattering amplitudes with the Luescher formalism to extract resonance parameters, including couplings to a variety of decay channels. After converting from SU(3) flavor states to physical flavor states, these couplings are extrapolated to their physical values by assuming that they only depend on an angular momentum factor, $k^\ell$, appropriate to the channel under consideration. This procedure gives 

\begin{align}
\Gamma(\pi_1) \to  b_1\pi &= 139-529\ \textrm{MeV} \nonumber \\ 
                 f_1 \pi  &= 0-24\ \textrm{MeV} \nonumber \\
                 \eta'\pi  &= 0-12\ \textrm{MeV}\nonumber  \\
                K^*\bar{K} &= 0-2\ \textrm{MeV} \nonumber \\
                 \eta \pi   &= 0-1\ \textrm{MeV}.
\end{align}

\begin{figure}[ht]
\includegraphics[width=12cm,angle=0]{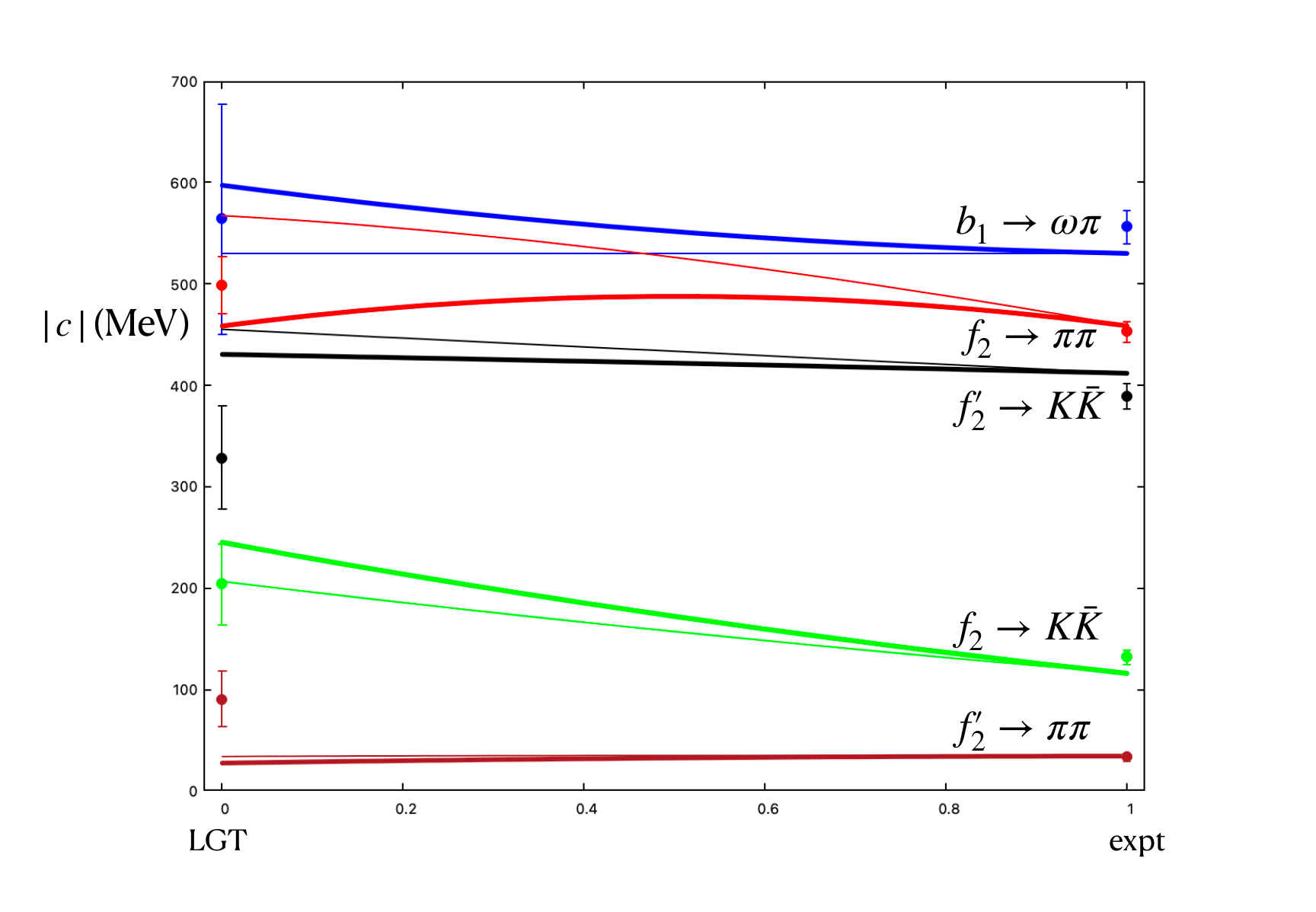}
\caption{Strong Decay Couplings. Experimental couplings are to the right; lattice computations at the $SU(3)$ point are to the left. Thick lines indicate extrapolations from 3P0 model couplings to the $SU(3)$ limit with form factors. Thin lines extrapolate without form factors.} 
\label{fig:b1-extrap}
\end{figure}

A number of model assumptions have been made in converting an \textit{ab initio} computation into phenomenologically useful predictions. For example, one expects a form factor to contribute to the extrapolation of the couplings from the SU(3) to the physical point. This can be tested (also within a model) by computing the form factors associated with the decays of ordinary mesons and comparing to the results using the lattice approach. In this regard,  Woss \textit{et al.} report couplings for a variety of canonical meson decays at the SU(3) point. In Fig. \ref{fig:b1-extrap} we show these couplings (labelled ``LGT") as points with error bars along with experimental couplings (labelled ``expt"). The lines show the scaling between these points as computed in the ``3P0'' model of hadronic couplings\cite{3p0}, with and without form factors. We see that the 3P0 model does a good job in reproducing the experimental points and that both extrapolations agree quite well with the lattice couplings, lending support to the lattice extrapolation.

Of course it is possible that extrapolations for hybrid mesons are not so successful. To avoid this, we compute hybrid decay couplings at the quark model analogue of the SU(3) point  and compare directly to the lattice couplings.
The quark model SU(3) point is obtained by fitting $s\bar s$ meson masses and using these model parameters to compute decay amplitudes. Using lattice SU(3) meson masses gives the couplings reported in Table \ref{tab:pi1}. Flavor SU(3) channels are listed in the first column, while the last column gives an idea of the range of predictions made in our model as obtained by performing a variety of fits under differing conditions. The results agree quite well with those obtained in the lattice calculation\cite{Woss:2020ayi}. Within the limited scope of this comparison, we conclude that the model is working well.

\begin{table}[ht]
\begin{tabular}{l|lcl}
\hline\hline
mode & LGT & preferred & model range \\
\hline
$\omega_8 \eta_8$ &  $0\to 279$ & 57 & $0 \to 82$ \\
$h_8 \eta_8$ & $978 \to 1909$ & 1230 & $780 \to 1450$ \\
$f_8 \eta_8$ & $0 \to 978$ & 920 & $580 \to 1070$ \\
\hline\hline
\end{tabular}
\caption{Lattice and model $\pi_1$ decay coupling amplitudes (MeV) at the $SU(3)$ point.}
\label{tab:pi1}
\end{table}

With these results, the simple lattice extrapolation model can be tested. In contrast to the case of ordinary mesons discussed above, we find quite large form factor effects, leading to approximately a factor of two reduction in the couplings at the physical point. 
We have therefore adopted two strategies in the following:  hybrid decay couplings are computed at the quark model analogue of the SU(3) point and extrapolated to the physical point in the same way as Woss \textit{et al.};  decay widths are computed directly at the ``physical" point.  Although the first method appears indirect, we remark that computing quark model matrix elements is ambiguous in situations with substantial relativistic effects. These can be ameliorated by computing in the nonrelativistic limit, extracting a matrix element, and then using Lorentz covariant dynamics to extrapolate to the physical limit\cite{Hayne:1981zy}. The first strategy emulates this procedure in a simple way, and coincides with that followed by Woss \textit{et al.}\cite{Woss:2020ayi}.


\section{Light Hybrid Decays and Discussion}
\label{sect:dec}

Results for partial decay widths for the hybrids in the $H_1$ multiplet are reported in this section. This multiplet is chosen because it contains the four lightest hybrids that are most likely to be detected in the near future. As shown in Table \ref{tab:JPC}, $H_1$ comprises the spin singlet $J^{PC}= 1^{--}$ vector, and the set of spin triplet hybrids with $J^{PC}= (0,1,2)^{-+}$. Notice that this includes the exotic $\pi_1$ state. Thus we examine the decay modes of 12 isoscalar and isovector hybrid mesons.

As discussed, we have generated an ``ensemble" of results by varying model assumptions in reasonable ways. We have also varied the assumed hybrid mass since it is an additional source of uncertainty. Clearly this is not a rigorous statistical exercise, but we feel that it nonetheless reveals useful information about model sensitivity.   
Hybrid masses have been obtained by making a rough extrapolation of naive lattice hybrid masses in the pion mass to the physical point. (``Naive" because the lattice masses are obtained from correlator plateaus, rather than from the Luescher formalism, which accounts for the resonant behavior of these states.) An additional problem is that lattice masses disagree by as much as 100 MeV, presumably because of ambiguity in scale setting (compare Refs. \cite{Dudek:2011bn} and \cite{Dudek:2013yja}). Lastly, this rough extrapolation gives a $\pi_1$ mass that is approximately 190 MeV high with respect to the (also approximate) experimental mass of 1600 MeV. We therefore shift all extrapolated masses downwards by 190 MeV to give the masses reported in the tables that follow. This procedure certainly does not yield a high fidelity hybrid spectrum! We look forward to a future in which experiment, lattice field theory, and models converge to a fuller understanding of the hybrid spectrum.

\subsection{Isovectors}
\label{sect:isovectors}

 Tables \ref{tab:LightIsovectorDecays} and \ref{tab:LightIsovectorDecays2} report partial widths for  the main decay modes for the isovector hybrids considered here.  The assumed hybrid mass is shown in the first column.  Decays modes are listed in the second column. The third column reports the central value of our results; the fourth column shows the range of results as model parameters are changed, the fifth column shows the range as the hybrid mass is changed by $\pm 100$ MeV, while the last column gives a range of values for the decay widths determined with the ``SU(3)" method.

We make some remarks on the results here. First, it is clear that decays to combinations of S-wave and P-wave mesons dominate the total widths. This agrees with ``S+P" rules found in many previous hybrid decay models\cite{
Page:1996rj} . We also find a moderate amount of model sensitivity to parameters, with roughly 50\% variations possible in the predicted partial widths, as shown in Table \ref{tab:LightIsovectorDecays}.

Widths obtained by extrapolating from the ``SU(3)" flavor point agree reasonably well with those obtained by direct computation, with the exception of the $J^{PC}= 1^{-+}$ hybrid, where the extrapolated widths are roughly twice as large as those obtained with physical parameters.

\begin{figure}
    \centering
    \includegraphics[width=\textwidth]{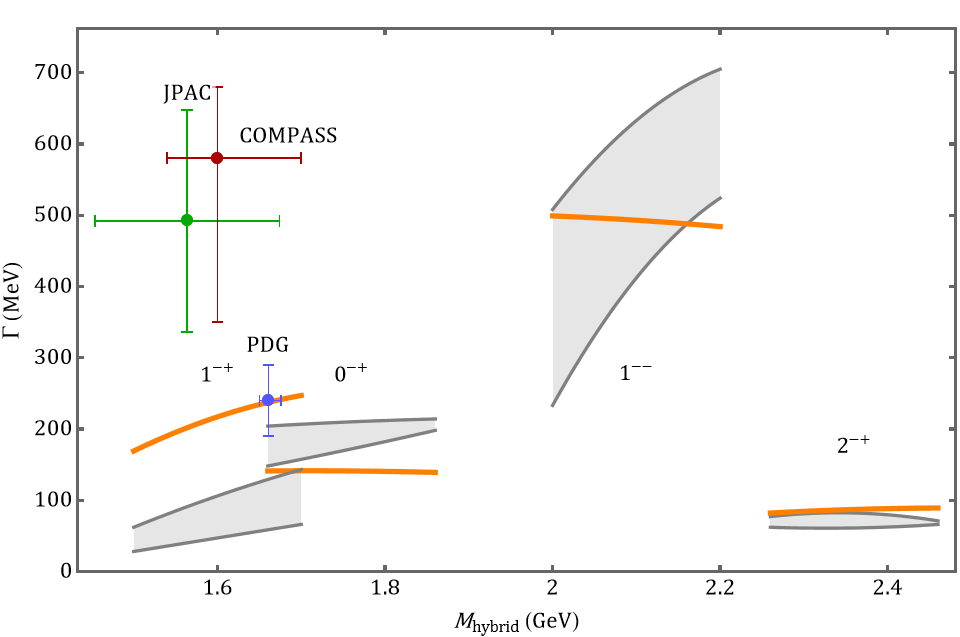}
    \caption{Isovector Decay Widths as Functions of the Hybrid Masses. The three points correspond to the PDG\cite{Workman:2022ynf}, COMPASS\cite{COMPASS:2018uzl} and JPAC\cite{COMPASS:2021ogp} values for the $1^{-+}$ width.}
    \label{fig:GlobalIsoDecayRanges}
\end{figure}

The situation is summarized in Fig. \ref{fig:GlobalIsoDecayRanges}, with grey regions illustrating deviations as model parameters and hybrid masses are changed, and orange lines showing similar dependence on the hybrid mass in the ``SU(3)" method. 
For the $1^{-+}$ we also report experimental results for the $\pi_1(1600)$ from the PDG, and for two analyses carried out by the COMPASS \cite{COMPASS:2018uzl} and JPAC \cite{COMPASS:2021ogp} collaborations. We note that the SU(3) method agrees very well with the total width reported in the PDG, but is approximately a factor of two below results from the more recent analyses. Of course, once resonances get so broad one can justifiably question the perturbative computation made here. If these results are confirmed it would appear that the perturbative result should be replaced by a resonance pole location obtained from a coupled channel analysis of the bare $\pi_1$ hybrid and its strongest decay channels.

\begin{table}[H]
    \centering
    \begin{tabular}{c|c|cccc}
Hybrid & Decay Channel & $\Gamma$ & $\Gamma(\alpha,\beta,\beta_m)$ & $\Gamma(M_{hyb})$ & $\Gamma_{SU(3)}$\\
\hline\hline
$~1^{--}~$ & $\eta\pi$ & - & - & - & - \\   
$(2100)$ & $\omega\pi$ &  0.9 & & \\
         & $\eta'\pi$ & - & - & - & -\\
         & $K^*\bar{K}$ & 0.5 & \\
         & $\rho f_0(500)$ & 40 & (34,55) & (44,36) & (35,\textbf{35},34)\\
         & $h_1(1170)\pi$ & - & - & - & -\\
         & $\rho\eta$ &  1.6 & \\
         & $a_1(1260)\pi$ & 74 & (72,180) & (77,69) & (263,\textbf{261},256)\\
         & $a_2(1320)\pi$ & 1.6 & \\
         & $\omega(1420)\pi$ & 0.4 & \\
         & $\rho\eta'$ & 1 & \\
         & $\rho f_0(980)$ & 62 & (50,72) & (61,59) & (77,\textbf{76},74)\\
         & $b_1(1235)\eta$ & - & - & - & - \\
         & $K_{1L}\bar{K}$ & 24 & (20,34) & (21,25) & (37,\textbf{37},36)\\
         & $K_{1H}\bar{K}$ & 45 & (33,68) & (33,51) & (87,\textbf{85},84)\\
         & $K^*(1410)\bar{K}$ & 0.1 & \\
         & $K_2^*(1430)\bar{K}$ & 1 & \\
         & $h_1(1170)\rho$ & 84 & (54,99) & (61,88) & (\o,\textbf{\o},\o)\\
         & $a_1(1260)\rho$ & 65 & (37,84) & (\o,75) & (\o,\textbf{\o},\o)\\
         & $b_1(1235)\rho$ & 150 & (93,177) & (\o,177) & (\o,\textbf{\o},\o)\\
         & $a_2(1320)\rho$ & 35 & (18,35) & (\o,114) & (\o,\textbf{\o},\o) \\
         \hline
$\Gamma_{tot}$ & & \textbf{585} & (421,802) & (632,701) & (499,\textbf{493},484) \\           \hline
$0^{-+}$ & $f_0(500)\pi$ & 51 & (51,146) & (56,46) & (52,\textbf{50},48)\\
$(1760)$ & $\rho\pi$ &  8.8 &  & \\
         & $f_0(980)\pi$ & 73.4 & (51,163) & (73,71) & (82,\textbf{83},82)\\
         & $K^*\bar{K}$ & 0.4 & \\         
         & $b_1(1235)\pi$ & 57 & (8,57) & (32,87) & (\o,\textbf{\o},\o)\\

\end{tabular}
\caption{Decay Modes for the Lightest Isovector Hybrids. x = negligible, - = forbidden by selection rules, \o= above threshold. The parameters used are obtained from a global fit to the \emph{uds} spectrum: $m_n=0.429$ GeV, $m_s=0.606$ Gev, $\alpha_S=1.496$, $\theta_{K}=33^\circ$, $\alpha=0.25$, $\beta=0.48$.}
    \label{tab:LightIsovectorDecays}
\end{table}

\begin{table}[H]
    \centering
    \begin{tabular}{c|c|cccc}
Hybrid & Decay Channel & $\Gamma$ & $\Gamma(\alpha,\beta,\beta_m)$ & $\Gamma(M_{hyb})$ & $\Gamma_{SU(3)}$\\
\hline\hline
         & $f_2(1270)\pi$ & x & \\
         & $f_0(1370)\pi$ & 22 & (4,31) & (14,28) & (7,\textbf{8},9)\\
         \hline
$\Gamma_{tot}$ & & \textbf{213} & (172,326) & (175,232) & 141,\textbf{141},139) \\            
\hline
$1^{-+}$ & $\eta\pi$ & - & - & - & -\\
$(1600)$ & $\rho\pi$ & 1.3 &  & & (2,\textbf{2},2)\\
         & $\eta'\pi$ & - & - & - & -\\
         & $b_1(1235)\pi$ & 54.5 & (39,92) & (34,72) & (142,\textbf{179},202)\\
         & $K^*\bar{K}$ & x & \\
         & $f_2(1270)\pi$ & x & \\
         & $f_1(1285)\pi$ & 9.7 & (6,17) & (5,14) & (25,\textbf{36},43)\\
         & $\rho\omega$ & - & - & - & -\\
\hline
$\Gamma_{tot}$ & & \textbf{65.8} & (46,106) & (39,86) & (169,\textbf{217},247)\\    
\hline
$2^{-+}$ & $f_0(500)\pi$ & 0.5 & \\
$(2360)$ & $\rho\pi$ & 3.7 & \\
         & $K^*\bar{K}$ & 0.8 & \\
         & $f_2(1270)\pi$ & 19 & (18,43) & (20,15) & (32,\textbf{34},36)\\
         & $f_1(1285)\pi$ & 0.6 & \\  
         & $f_0(980)\pi$ & 0.4 & \\
         & $f_0(1370)\pi$ & x & \\
         & $\rho(1450)\pi$ & 7.2 & \\
         & $K_{1L}\bar{K}$ & x & \\
         & $K_{1H}\bar{K}$ & 0.1 & \\
         & $a_1(1260)\eta$ & 0.2 & \\
         & $a_2(1320)\eta$ & 12 & (10,23) & (13,11) & (15,\textbf{16},16)\\ 
         & $K^*(1410)\bar{K}$ & 0.3 & \\
         & $K_0^*(1430)\bar{K}$ & x & \\
         & $K_2^*(1430)\bar{K}$ & 39 & (33,66) & (37,39) & (32,\textbf{34},36)\\
         & $a_0(1450)\eta$ & x & \\
         \hline
$\Gamma_{tot}$ & & \textbf{83.8} & (61,132) & (84,82) & (82,\textbf{87},89)\\    
\hline\hline
    \end{tabular}
    \caption{Continuation of Table \ref{tab:LightIsovectorDecays} above.}
    \label{tab:LightIsovectorDecays2}
\end{table}

\subsection{Isoscalars}

We consider isoscalar hybrids to be linear combinations of light and strange quarks defined by

\begin{eqnarray}
    \ket{L}&=&\cos{\theta_{I}}\ket{n\bar{n}}-\sin{\theta_{I}}\ket{s\bar{s}} \nonumber \\
    \ket{H}&=&\sin{\theta_{I}}\ket{n\bar{n}}+\cos{\theta_{I}}\ket{s\bar{s}},
\end{eqnarray}
where $\theta_{I}$ is the mixing angle. Of course, nothing is known about hybrid mixing angles experimentally. An ambitious lattice computation does, however, report light hybrid mixing angles\cite{Dudek:2013yja}. We regard these computations as preliminary and therefore choose to present results as a function of the mixing angle. The exception to this are the vector isoscalar states, which are expected to be ideally mixed by quite general arguments\cite{Swanson:2023zlm}, and therefore $\theta_I(1^{--})$ is taken to be zero.

Decay widths for eight isoscalar mesons in the $H_1$ multiplet are presented in Table \ref{tab:LightIsoscalarDecays} as a function of mixing angle. The assumed masses for the isoscalar hybrids are reported in column one under the hybrid $J^{PC}$ and  are estimated as described in Section
\ref{sect:dec}.

\begin{figure}[h!]
    \centering
    \includegraphics[width=0.45\textwidth]{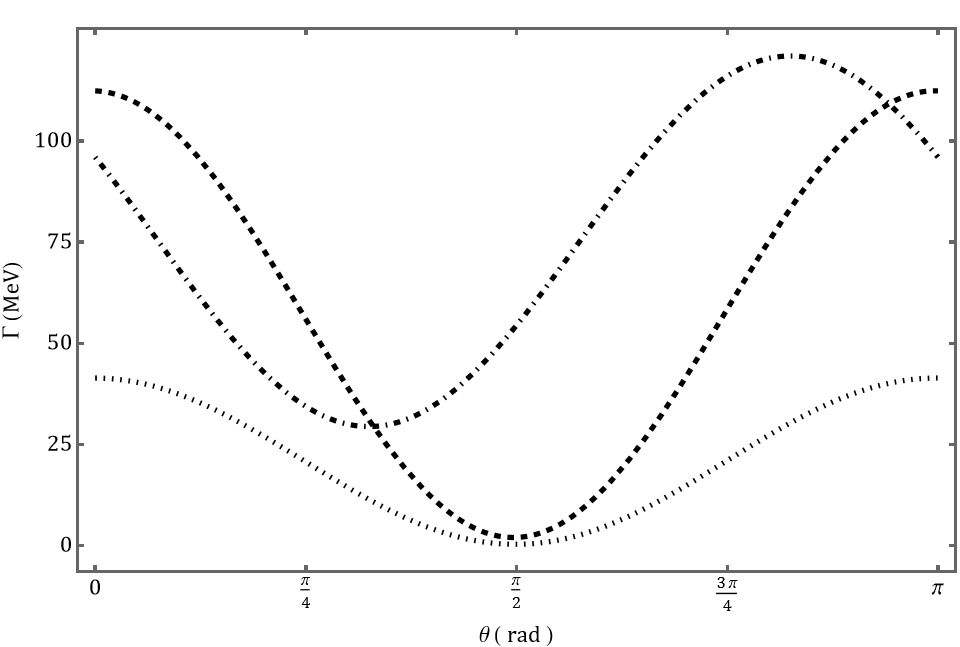}
    \includegraphics[width=0.45\textwidth]{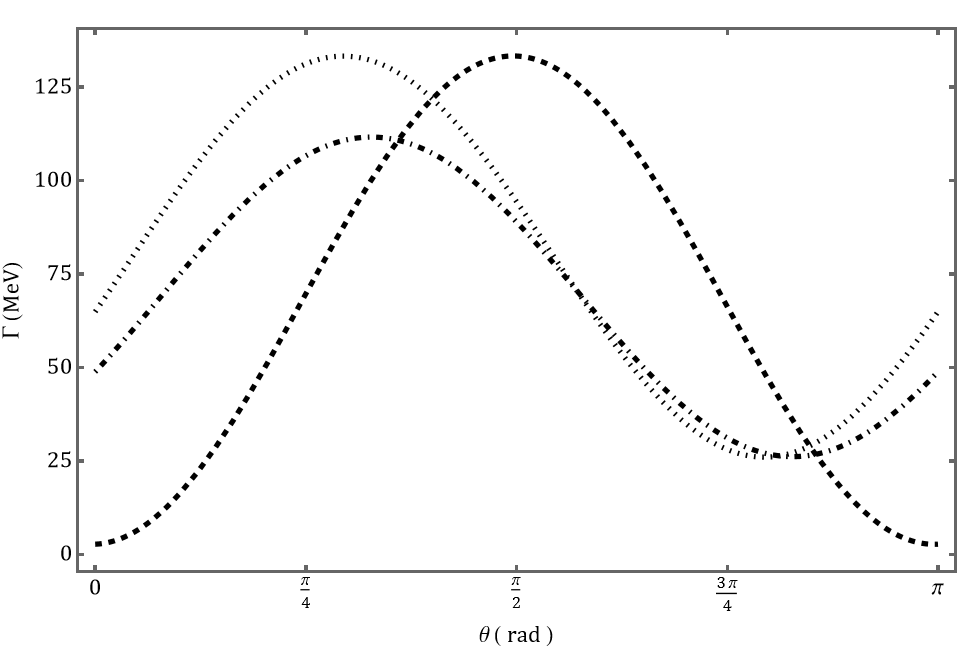}
    \caption{Isoscalar Total Widths as Functions of the Mixing Angle, $\theta_{I}$. Dashed line = $0^{-+}$, dotted line = $1^{-+}$, dot-dashed line = $2^{-+}$.}
    \label{fig:isoscalarsLowHighWidths}
\end{figure}

We find that the ideally mixed vector widths are approximately 90 MeV. Other widths depend on the mixing angle assumed. As shown in Fig. \ref{fig:isoscalarsLowHighWidths}, the low mass states have minimal widths ranging from a few MeV to 30 MeV. 
The possible low widths for the light isoscalar $0^{-+}$ and $1^{-+}$ arise because these states are estimated to have masses of 1750 and 1650  MeV respectively, and therefore are below threshold for decay to S+P kaons, yielding total widths that are approximately proportional to $\cos^2(\theta_I)$. Lattice calculations indicate that mixing angles tend to be small\cite{Woss:2020ayi}, so we judge this scenario to be unlikely. It is intriguing that the high mass isoscalar $0^{-+}$ is expected to be quite narrow under these assumptions (the $0^{-+}$ mixing angle is approximately 20 degrees, giving a total width of 18 MeV for the $s\bar{s}g$ hybrid).

\begin{table}[H]
\centering
\scalebox{0.75}{
\begin{tabular}{c|c|c|cl}
Hybrid $J^{PC}$ & Decay Channel & $\Gamma_L$ & $\Gamma_H$ \\
\hline
\hline
H1 states \\
\hline
$1^{--}$ & $\rho\pi$ & 10 & - \\
$(2120-2250)$ & $\omega\eta$ & 0.2 & - \\
         & $K^*\bar{K}$ &  1.1 & 2.6 \\
         & $\rho(1450)\pi$ & 2.1 & - \\
         & $\omega\eta'$ & 0.1 & - \\
         & $\phi\eta$ & - & x \\
         & $\phi\eta'$ & - & x \\
         & $K_{1L}\bar{K}$ & 24 & 28.8 \\
         & $K_{1H}\bar{K}$ & 46 & 61.7 \\
         & $K^*(1410)\bar{K}$ & 0.2 & 1.3 \\
         & $K_2^*(1430)\bar{K}$ & x & 0.5 \\
         & $\omega(1420)\eta$ & x & - \\
         \hline
$\Gamma_{tot}$ & & 84.5 & 95 \\
\hline
$0^{-+}$ & $f_0(500)\eta$ & $33.3\cos^2{\theta}$ & $30.2\sin^2{\theta}$ \\
$(1750-1850)$ & $K^*\bar{K}$ & $(\cos{\theta}\sqrt{0.8} - \sin{\theta}\sqrt{2})^2$ & $(\sin{\theta}\sqrt{1.1} + \cos{\theta}\sqrt{2.7})^2$ \\
         & $a_2(1320)\pi$ & $1.0\cos^2{\theta}$ & $2.5\sin^2{\theta}$ \\
         & $f_0(980)\eta$ & $42.8\cos^2{\theta}$ & $44.9\sin^2{\theta}$ \\
         & $a_0(1450)\pi$ & $34.5\cos^2{\theta}$ & $54.6\sin^2{\theta}$ \\
         & $\phi\eta$ & x & x \\
         & $f_2(1270)\eta$ & \o & x \\
         & $\phi\eta'$ & \o & \o \\
         & $K^*(1410)\bar{K}$ & \o & \o \\
         & $f_0(1370)\eta$ &  \o & \o \\
         & $K_0^*(1430)\bar{K}$ & \o & \o \\
         & $K_2^*(1430)\bar{K}$ & \o & \o \\
         & $f_2'(1525)\eta$ & \o & \o \\
\hline
$\Gamma_{tot}$ & & $112\cos^2(\theta) - 2.5\cos(\theta)\sin(\theta) + 2\sin^2(\theta)$ & $2.7\cos^2(\theta) + 3.4\cos(\theta) \sin(\theta) + 
 133\sin^2(\theta)$ \\
\hline
$1^{-+}$ & $\eta\eta'$ & - & - \\
$(1650-1800)$ & $K^*\bar{K}$ & $(\cos{\theta}\sqrt{0.1} - \sin{\theta}\sqrt{0.3})^2$ & $(\sin{\theta}\sqrt{0.22} + \cos{\theta}\sqrt{0.6})^2$\\
         & $a_1(1260)\pi$ & $41.3\cos^2{\theta}$ & $53.2\sin^2{\theta}$ \\
         & $f_2(1270)\eta$ & \o & \o \\
         & $f_1(1285)\eta$ & \o & \o \\
         & $\pi(1300)\pi$ & - & - \\
         & $a_2(1320)\pi$ & x & $.15\sin^2{\theta}$ \\
         & $K_{1L}\bar{K}$ & \o & $(\sin{\theta}\sqrt{40.8} + \cos{\theta}\sqrt{64.3})^2$ \\
\end{tabular}}
\caption{Decay modes for the lightest isoscalar hybrids. x = negligible, - = forbidden by selection rules, \o = above threshold. The parameters used are obtained from a global fit to the \emph{uds} spectrum.}
\label{tab:LightIsoscalarDecays}
\end{table}

\begin{table}[H]
    \centering
\scalebox{0.80}{
\begin{tabular}{c|c|c|cl}
Hybrid $J^{PC}$ & Decay Channel & $\Gamma_L$ & $\Gamma_H$ \\
\hline
\hline
         & $K_{1H}\bar{K}$ & \o & \o \\ 
         & $K^*(1410)\bar{K}$ & \o & \o \\
         & $K_2^*(1430)\bar{K}$ & \o & \o \\
         & $\phi\eta$ & x & x \\
         & $\phi\eta'$ & \o & \o \\
         & $f_2'(1525)\eta$ & \o & \o \\
\hline
$\Gamma_{tot}$ & & $41.4\cos^2(\theta) - 0.34\cos(\theta)\sin(\theta) + 0.3\sin^2(\theta)$ & $64.9\cos^2(\theta) + 103\cos(\theta)\sin(\theta) + 
 94.4\sin^2(\theta)$ \\
\hline
$2^{-+}$ & $K^*\bar{K}$ & $(\cos{\theta}\sqrt{0.87} - \sin{\theta}\sqrt{1.4})^2$ & $(\sin{\theta}\sqrt{0.94} + \cos{\theta}\sqrt{1.38})^2$ \\
$(2430-2560)$ & $\rho\pi$ & $5.5\cos^2{\theta}$ & $5.4\sin^2{\theta}$ \\ 
         & $a_2(1320)\pi$ & $41\cos^2{\theta}$ & $32.9\sin^2{\theta}$\\
         & $a_1(1260)\pi$ & $2.4\cos^2{\theta}$ & $2.9\sin^2{\theta}$ \\  
         & $a_0(1450)\pi$ & $0.26\cos^2{\theta}$ & $0.38\sin^2{\theta}$ \\
         & $\rho(1450)\pi$ & $2.4\cos^2{\theta}$ & $3\sin^2{\theta}$ \\
         & $K_{1L}\bar{K}$ & $(\cos{\theta}\sqrt{0.1} - \sin{\theta}\sqrt{0.17})^2$ & $(\sin{\theta}\sqrt{0.15} + \cos{\theta}\sqrt{0.17})^2$\\
         & $f_2(1270)\eta$ & $6.0\cos^2{\theta}$ & $5.0\sin^2{\theta}$ \\
         & $K_{1H}\bar{K}$ & $(\cos{\theta}\sqrt{0.18} - \sin{\theta}\sqrt{0.29})^2$ & $(\sin{\theta}\sqrt{0.29} + \cos{\theta}\sqrt{0.44})^2$\\  
         & $K^*(1410)\bar{K}$ & $(\cos{\theta}\sqrt{0.4} - \sin{\theta}\sqrt{1.1})^2$ & $(\sin{\theta}\sqrt{0.57} + \cos{\theta}\sqrt{1.27})^2$ \\
         & $K_0^*(1430)\bar{K}$ & $(\cos{\theta}\sqrt{x} - \sin{\theta}\sqrt{0.17})^2$ & $(\sin{\theta}\sqrt{x} + \cos{\theta}\sqrt{0.26})^2$ \\
         & $K_2^*(1430)\bar{K}$ & $(\cos{\theta}\sqrt{39} - \sin{\theta}\sqrt{39.3})^2$ & $(\sin{\theta}\sqrt{37.3} + \cos{\theta}\sqrt{33.3})^2$\\  
         & $\phi\eta$ & $0.14\sin^2{\theta}$ & x \\
         & $\phi\eta'$ & $0.10\sin^2{\theta}$ & x \\
         & $f_2'(1525)\eta$ & $13\sin^2{\theta}$ & $12\cos^2{\theta}$ \\
\hline
$\Gamma_{tot}$ & & $97.9\cos^2(\theta) - 82\cos(\theta)\sin(\theta) + 
 53.2\sin^2(\theta)$ & $49\cos^2(\theta) + 75\cos(\theta)\sin(\theta) + 
 89.4\sin^2(\theta)$ \\
\hline\hline
    \end{tabular}}
    \caption{Continuation of Table \ref{tab:LightIsoscalarDecays} above.}
    \label{tab:LightIsoscalarDecays2}
\end{table}

\begin{figure}[h!]
    \centering
    \includegraphics[width=0.45\textwidth]{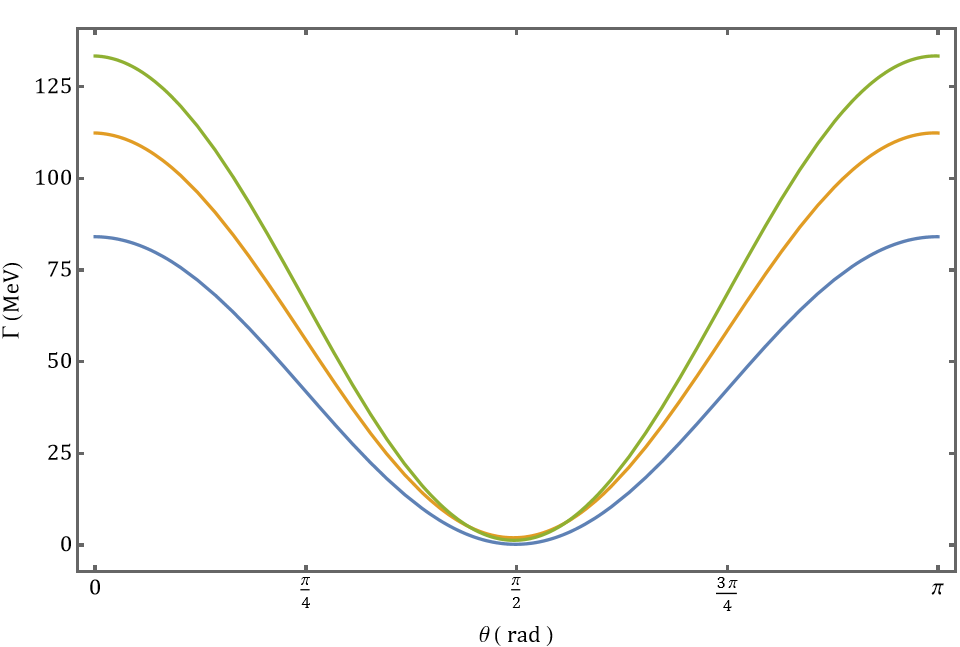}
    \includegraphics[width=0.45\textwidth]{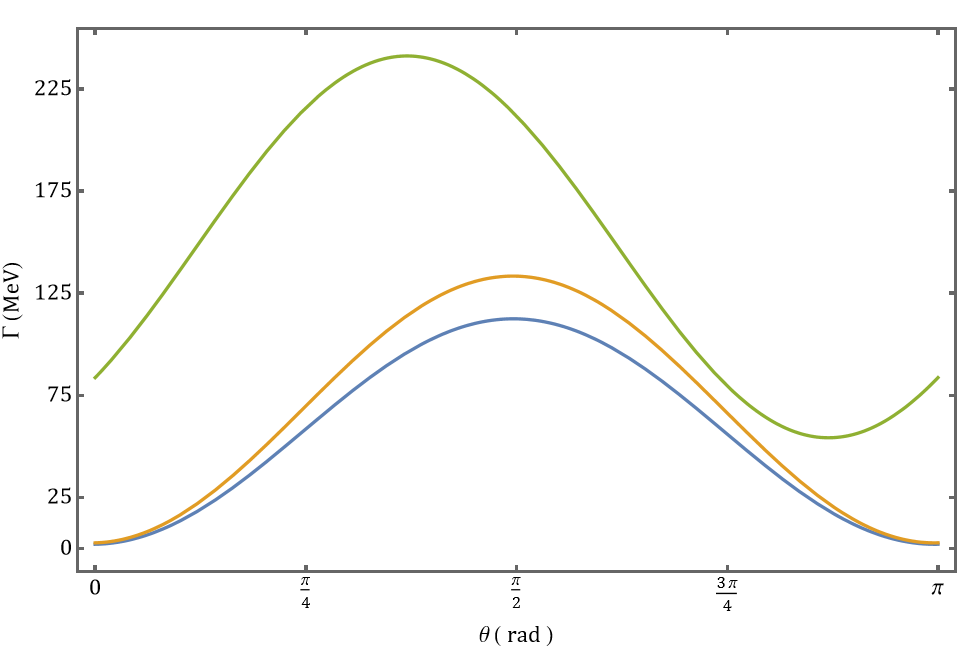}
    \caption{Isoscalar $0^{-+}$ total widths as functions of the mixing angle, $\theta_{I}$, for three different values of the hybrid mass: 1650, 1750, 1850 MeV for the low state, 1750, 1850, 1950 Mev for the high state. Blue = 1650 (1750) MeV, orange = 1750 (1850) MeV, green = 1850 (1950) MeV.}
    \label{fig:0-+WidthsVsMass}
\end{figure}

Fig. \ref{fig:0-+WidthsVsMass} displays the dependence of the total widths on the assumed $0^{-+}$ isoscalar hybrid masses. Variations on the order of 20\% are evident at small mixing angles, with a large increase in the $s\bar{s}g$ width at larger mass due to a substantial $K^*(1410)\bar K$ mode that goes on-shell.

\begin{figure}[h!]
    \centering
    \includegraphics[width=0.45\textwidth]{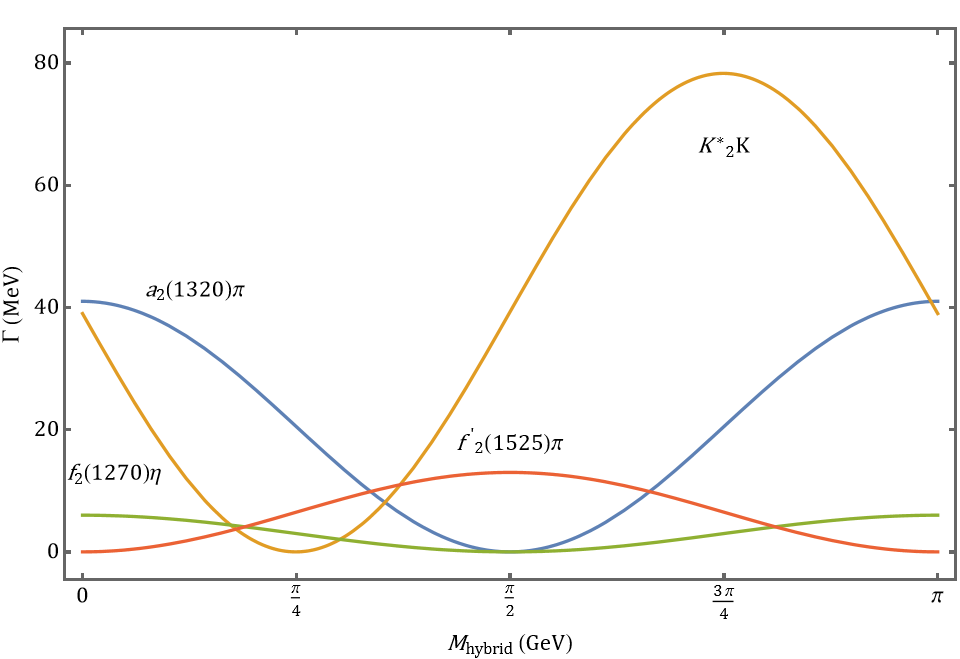}
    \includegraphics[width=0.45\textwidth]{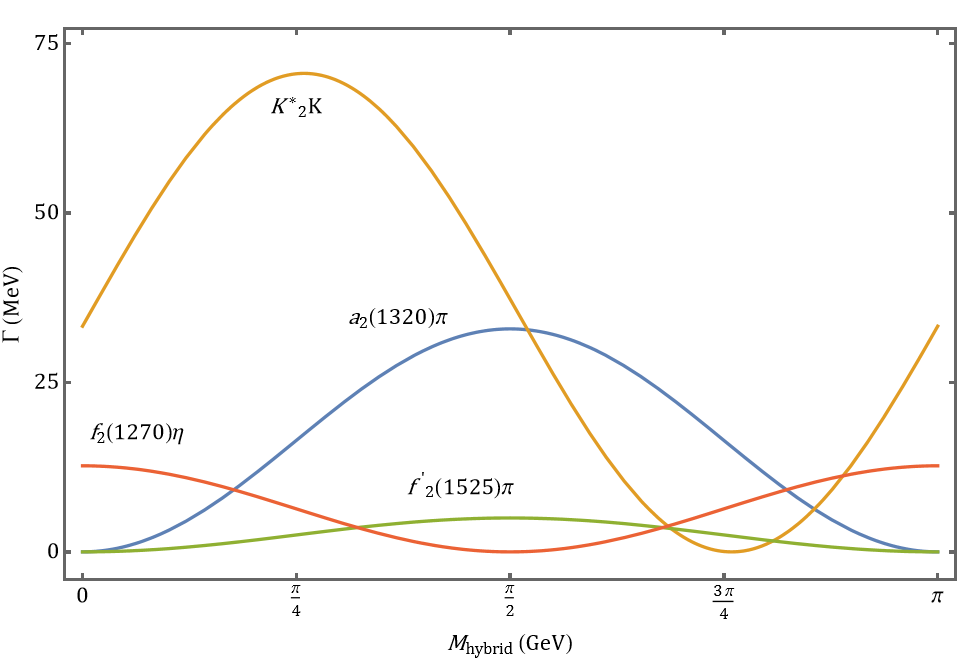}
    \caption{Largest channels for the low (left) and high (right) $2^{-+}$ isoscalar hybrids as functions of the mixing angle.}
    \label{fig:2-+Widths}
\end{figure}

Similarly, Fig. \ref{fig:2-+Widths} displays the largest decay modes of the $2^{-+}$ hybrid as a function of  mixing angle. The rapid variation of some of the channels implies that it might be possible to constrain the relevant mixing angles if enough partial widths are measured. 

\begin{figure}[h!]
    \centering
    \includegraphics[width=0.8\textwidth]{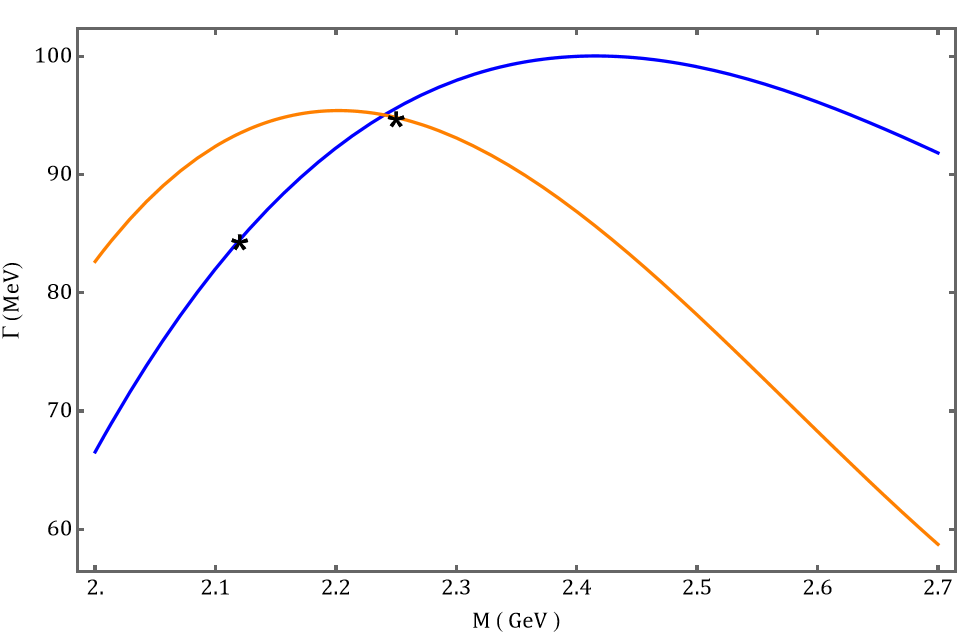}
    \caption{Isoscalar $1^{--}$ total decay widths as function of the hybrid mass. Blue: light quark state, orange: strange quark state. Values at the nominal masses are denoted with a star.}
    \label{fig:1--IsoscalarTotWidths}
\end{figure}

Determining model sensitivity is more involved than the isovector case because of the mixing angle dependence of the partial widths. In this regard,  Fig.  \ref{fig:1--IsoscalarTotWidths} shows the dependence of the isoscalar vector total decay widths as the assumed hybrid mass is varied. Changes of approximately 10-20\% are seen, as with the isovectors. 

Fig. \ref{fig:isoscalarsLowHighWidths} shows the total widths for the light and heavy $J^{PC}=2^{-+}$ isoscalar hybrids as a function of assumed mass and mixing angle. Again, the mass-dependence is not strong, while the angle-dependence can be used as a possible way to measure this parameter.

\begin{figure}[h!]
    \centering
    \includegraphics[width=\textwidth]{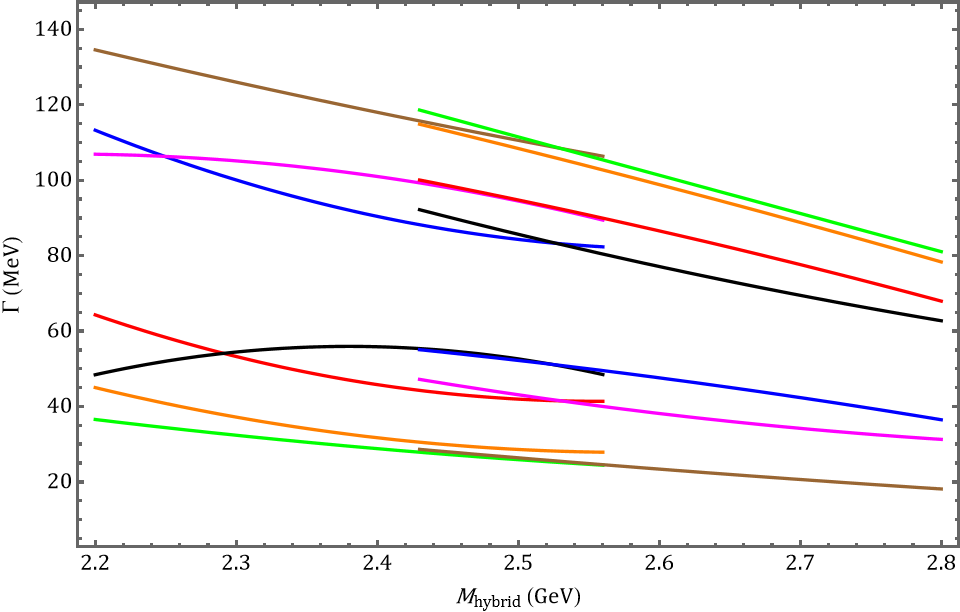}
    \caption{Low and high isoscalar $2^{-+}$ total widths as functions of the hybrid mass, for various values of the mixing angle $\theta_{I}$. Blue: $\theta_{I}=0=\pi$, red: $\theta_{I}=\pi/6$, orange: $\theta_{I}=\pi/4$, green: $\theta_{I}=\pi/3$, black: $\theta_{I}=\pi/2$, magenta: $\theta_{I}=2\pi/3$, brown: $\theta_{I}=5\pi/6$}.
    \label{fig:2-+isoscalarsTotWidths}
\end{figure}

\section{Discussion and Conclusions}
\label{sect:conc}

The phenomenological implications of this study are presented in this section. We argue that three good hybrid meson candidates are known and suggest strategies to assist in confirming that they are exotic and in searching for possible partner states.

\subsection{$H_1(1^{--})$}

The results shown in Table \ref{tab:LightIsovectorDecays} imply that it may be very difficult to observe an isolated isovector vector hybrid meson in reactions like $e^{+}e^{-} \to H_1(1^{--}) \to $ hadrons, since the state is so broad (the decay constant for this reaction is discussed in Ref. \cite{Swanson:2023zlm}). Thus direct observation of this state may be impossible; however, if data are collected on sufficiently many channels, it may be possible to observe the hybrid state as a resonance pole  obtained from a unitary analysis such as that of Ref. \cite{Husken:2022yik}. Alternatively, the isoscalar vector hybrids are expected to have moderate total widths around 90 MeV.  This is encouraging and suggests that measurements of $K_1\bar{K}$ production can be very helpful in hybrid searches.

\subsection{$H_1(0^{-+})$}

The $\pi(1800)$ has been suggested as a possible exotic meson\cite{pdg-pi} because it preferentially decays to S+P modes, as expected for hybrid mesons.  This suggestion is examined here.

Experimental and theoretical pseudoscalar masses are listed in Table \ref{tab:pi}. The first row summarizes observed states; the second gives the plateau masses as determined at a pion mass of 391 MeV\cite{Dudek:2013yja}, the final rows are model computations of canonical pseudoscalar masses. There is general agreement in all approaches concerning the $\pi(1300)$, lending support to the identification of this state as the quark model $2{}^1S_0$. Given this, the $3{}^1S_0$ is expected in the range 1700-1900 MeV, which aligns nicely with the $\pi(1800)$. However, this state is not seen in $K^*\bar K$ or $\rho\pi$, and is seen in S+P modes such as $f_0\pi$ and $K_0^*\bar K$. Furthermore, the ``estimated" $H_1(0^{-+})$ hybrid mass is 1760 MeV -- quite close to the observed mass.

\begin{table}[h]
\begin{tabular}{l|cccc|cccccc}
\hline\hline
source & \multicolumn{4}{c}{$\pi$ masses} &  \multicolumn{2}{c}{$\eta$ masses} \\
\hline
RPP\cite{Workman:2022ynf} & 139 & 1300 & 1800 & -- & 548 & 958 & 1294 & 1409 & 1475 & 2221 \\
LGT\cite{Dudek:2013yja}  &   391 & 1520 & 2100\footnote{gluonic content} & 2350   & 600 & 1000 & 1680 & 1750 & 2100$^a$ & 2300$^a$ \\
GI\cite{Godfrey:1985xj}   &   150 & 1300 & 1880 &  & 520 & 960 & 1440 & 1630 &  &  \\
fit A  & 560 & 1365 & 1720 & 1990 & 560 & 730 & 1365 & 1580   & 1990 & 2200 \\
fit A'  & 600 & 1390 & 1750 & 2030 & 600 & 820 & 1390 & 1580   & 1750 & 1940 \\
\hline\hline
\end{tabular}
\caption{$0^{-+}$ Meson Masses (MeV).}
\label{tab:pi}
\end{table}

In view of this observation, it is useful to compare decay modes of canonical and hybrid mesons as predicted by the 3P0 model and this model. A variety of partial widths are presented in Table \ref{tab:piDecay}. For comparison, we show the predicted decays of the $\pi(1300)$, obtaining agreement with the (poorly known) PDG value. Similar computations for a $\pi(3S)$ at an assumed mass of 1850 MeV can be compared to  hybrid partial widths, as shown. The most striking difference between an excited pion and pseudoscalar hybrid is that the spin selection rule disallows the decay to $b_1\pi$ for canonical pions, making this a smoking gun mode for an exotic candidate. All the other modes also have large ratios between canonical and hybrid states, which should make identifying a hybrid $H_1(\pi)$ reasonably easy if partial widths can be measured.

\begin{table}[ht]
\begin{tabular} {l|cccccccc|ll}
\hline\hline
state &  $f_0(500)\pi$ & $\rho\pi$ & $f_0(980)\pi$ & $K^*\bar K$ & $b_1(1235)\pi$ & $f_2(1270)\pi$ &  $f_0(1370)\pi$ & $\omega\rho$ & $\Gamma_{tot}$ & $\Gamma_{PDG}$ \\
\hline
$\pi(2S)$ &  1 & 230 & 2 & \o & -- & \o & \o & \o & 233 & 200-600\\
$\pi(3S;1850)$ & 12 & 117 & x & 64 & -- & 30 & 1 & 109 & 333+ & 215 \\
$H_1(\pi)$ & 51 & 9 & 73 & 0 & 57 & x & 22 & x  & 212 & -- \\
\hline\hline
\end{tabular}
\caption{Predicted Decay Modes of $\pi$ States (MeV).}
\label{tab:piDecay}
\end{table}

If the $\pi(1800)$ is confirmed as a hybrid meson it is natural to search for a nearby $\pi(3S)$, which should have largely complementary decay modes. Intriguingly, the COMPASS collaboration sees evidence (which is not discussed) for two $\pi(1800)$ states in the $[\pi\pi]_S\pi$ channel of an isobar analysis of   $\pi p \to 3\pi p$\cite{COMPASS:2015gxz} (Fig. 24), with the enhancements near 1750 and 1850 MeV. We remark that the approximate orthogonality of the decay channels seen in Table \ref{tab:piDecay} implies that the  $\pi(3S)$ and $H_1(\pi)$ may not move much from their bare values, supporting the claim that two nearby states can be found near 1800 MeV.
If this scenario is correct then it is natural to seek hybrid isoscalar partner
states  near 1800 and 1900 MeV. Of these, the heavy $s\bar{s}g$ state is predicted to be anomalously narrow (10's of MeV) and might therefore be an attractive option for searches.

\subsection{$H_1(1^{-+})$}

We now turn attention to the exotic $J^{PC}= 1^{-+}$ states. The $\pi_1$ has been mentioned in Sect. \ref{sect:isovectors} as a plausible hybrid candidate. This state has a confused history, with claims of signals near 1400 and 1800 MeV\cite{Workman:2022ynf}. These claims have been challenged, with a consensus emerging that only one $\pi_1$ exists near 1600 MeV (results from two analyses are  $M= 1564 \pm 24\pm 86$ MeV, $\Gamma= 492 \pm 54 \pm 102$ MeV \cite{JPAC:2018zyd} and $M = 1623 \pm 47 \pm 50$ MeV, $\Gamma = 455 \pm 88 \pm 150$ MeV \cite{Kopf:2020yoa}).  As a result we have used this state as a benchmark in the previous discussion. As mentioned, our ``SU(3)" width agrees well with the width reported in the RPP, although both are too narrow with respect to the recent pole position determinations. Of course, such large widths cannot be reliably predicted in perturbation theory, and we regard this disagreement  as an indication that a nonperturbative calculation is required, rather than that the model is disfavored.

Important additional information has recently been obtained by the BESIII Collaboration, who have observed an isoscalar exotic state in $J/\psi \to \gamma \eta \eta'$, called the $\eta_1(1855)$, with mass 
$1855 \pm 9 {}^{+6}_{-1}$ MeV and width $188 \pm 18 {}^{+3}_{-8}$  MeV\cite{bes-eta1}. The mass of the state makes it an ideal candidate for the partner heavy isoscalar $1^{-+}$ hybrid\cite{Swanson:2023zlm}. In addition, the right panel of Fig. \ref{fig:isoscalarsLowHighWidths} indicates that the flavor mixing angle is expected to be approximately 10 (or 80) degrees. This is in remarkable (considering the uncertainties involved) agreement with the lattice determination of 21(5) degrees for this angle\cite{Dudek:2013yja}, providing further evidence for the identification of the $\eta_1(1855)$ as the ``$s\bar{s} g$'' analogue of the $\pi_1(1600)$. Given this observation, one expects a light isoscalar exotic with mass near 1750 MeV with a total width of approximately 150 MeV (Fig. \ref{fig:isoscalarsLowHighWidths}, left panel) and dominant decays to $a_1\pi$ and $K_1\bar{K}$.

We remark that similar observations have been made by Chen \textit{et al.}, who use adiabatic gluonic surfaces to model the hybrid spectrum and our model for hybrid decays\cite{Chen:2023ukh}. Additional interpretations have also been proposed that are based on the flux tube model\cite{Qiu:2022ktc} and model lagrangians\cite{Shastry:2023ths}.

%

\subsection{$H_1(2^{-+})$}

The $H_1(2^{-+})$ states  have  predicted widths of approximately 80 MeV for the isovector and, using a mixing angle of $\theta_I = 10(5)$ degrees\cite{Dudek:2013yja}, 75(8) MeV and 63(9) MeV for the isoscalar states. It is therefore possible that a $2^{-+}$ hybrid state is observable in the spectrum. Unfortunately, neither lattice computations nor modelling are capable of determining the spectrum accurately enough to permit identifying a hybrid by its mass alone. Table \ref{tab:pi2} shows all observed isovector and isoscalar states in the row labelled RPP. Results from a comprehensive lattice computation are also listed. Evidently the agreement with experiment is not good (the lattice computation is made at a pion mass of 391 MeV, but pion-mass dependence is weak in this channel). The row labelled GI refers to quark model results from Godfrey and Isgur; although the lowest multiplet agrees reasonably well with experiment, it appears that the $\pi_2(2S)$ mass is much higher than the likely experimental candidate at 1874 MeV. The final two rows are
predictions from nonrelativistic quark models as described in Sect.  \ref{sect:model} (see also Table \ref{tab:models2}). Again, the agreement with experiment is not encouraging. Thus it appears that present capabilities in meson mass predictions are not sufficient to suggest a hybrid meson candidate.

\begin{table}[h]
\begin{tabular}{l|cccc|cc}
\hline\hline
source & \multicolumn{4}{c}{$\pi_2$ masses} &  \multicolumn{2}{c}{$\eta_2$ masses} \\
\hline
RPP\cite{Workman:2022ynf} & 1670 & 1874 & 1963\footnote{not established} & 2090$^a$ & 1617 & 1842 \\
LGT\cite{Dudek:2013yja}  &   1900 & 2350\footnote{gluonic content} & 2550 & -- & 1900 & 2000\\
GI\cite{Godfrey:1985xj}   &   1680 & 2130 & -- & -- & 1680 & 1890 \\
fit A  & 1580 & 1880 & 2130 & 2350 & 1580 & 1785  \\
fit A'  & 1590 & 1865 & 2100 & 2300 & 1590 & 1820  \\
\hline\hline
\end{tabular}
\caption{$2^{-+}$ Meson Masses (MeV).}
\label{tab:pi2}
\end{table}

Another possibility for identifying a hybrid meson is via 
signatures in its decay modes. We examine this by computing canonical $\pi_2(nS)$ decays using the well-established 3P0 model of hadronic decay\cite{3p0}. To make concrete comparisons we assume that the $H_1(2^{-+})$ isovector hybrid has mass 2250 MeV and that there is also a 3S radial $q\bar q$ excitation at the same mass. Results are reported in Table \ref{tab:pi2Decay}. Note that the model does quite well in predicting the widths of the $\pi_2(1670)$ and $\pi_2(1880)$. Comparing the final two rows reveals no distinct features (given model uncertainties) that can be used to distinguish possible hybrid and canonical states. However, ratios, such as $f_2\pi/\rho\pi$, are more robust predictions of the models, and can provide useful diagnostics. In this case, $f_2\pi/\rho\pi \approx 1$ for a $q\bar q(3S)$ state, while it is approximately 5 for a hybrid.  Similarly, the S-D-G amplitude ratios for $\pi_2 \to f_2 \pi$ are approximately 1:0:0 for a hybrid but are 1 : 0.14 : 0.003 for the $q\bar q(1S)$ state, 1 : 0.62 : 0.06 for the $q\bar q(2S)$ state, and -0.24 : 1 : -0.38 for a $q\bar q(3S)$ state near 2250 MeV. 

The COMPASS collaboration finds strong evidence for a $\pi_2$ resonance near 1880 MeV\cite{COMPASS:2015gxz}. They note that the $\pi_2(1670)$ has a strong coupling to $f_2\pi$ in S-wave, while the $\pi_2(1880)$ couples dominantly in D-wave. This is consistent with the ratios just quoted, and  supports the identification of the $\pi_2(1670)$ as a canonical  2S state and the $\pi_2(1880)$ as a canonical 3S state. This conclusion is in opposition to previous suggestions that the $\pi_2(1880)$ may be a hybrid meson\cite{Barnes:1996ff,Close:1994hc}. This claim was based on old expectations of hybrid masses near 1800 MeV and the $2^{-+}$ spectrum of Ref. \cite{Godfrey:1985xj} (reported in row 4 of Table \ref{tab:pi2}. As discussed above, we believe that both of these assumptions are incorrect.

It is remarkable that the COMPASS collaboration also sees evidence for a $\pi_2$ near 2360 MeV.
The enhancement is seen in $(\pi\rho)_F$ (Fig. 21 of  Ref. \cite{COMPASS:2015gxz}), $(f_0(980)\pi)_D$, and $([\pi\pi]_S\pi)_D$ in Fig. 25. There is also a hint of a signal in $(\pi f_2)_S$ in Fig 22.   These signals are not discussed in the paper and we therefore strongly suggest that the data be analyzed for possible resonant behavior in the region of 2360 MeV. These signals are in accord with 
our expectations with the notable exception of the $(\pi \rho)_F$ mode, which we predict to be much smaller than the (unreported) $(\pi\rho)_P$ mode.

\begin{table}[ht]
\begin{tabular} {l|cccccc|ll}
\hline\hline
state &  $\pi\rho$ & $\omega\rho$ & $f_2\pi$ & $K^*\bar K$ & $\rho(1450)\pi$ & $K_2^*\bar{K}$ & $\Gamma_{tot}$ & $\Gamma_{PDG}$ \\
\hline
$\pi_2(1S)$ & 117 & 25 & 85 & 20 & 3 & -- & 250 & 258(8) \\
$\pi_2(2S)$ & 86 & 24 & 31 & 3 & 19 & 9 & 172+ & 237(30) \\
$\pi_2(3S;2250)$ & 12 & 10 & 16 & 6 & 35 & 8 & 87+ & -- \\
$H_1(\pi_2)$ & 4 & 0 & 20 & 0 & 7 & 40 & 71 & -- \\
\hline\hline
\end{tabular}
\caption{Predicted Decay Modes of $\pi_2$ States (MeV). The $\pi_2(1S)$ results coincide with those of Ref.\cite{Barnes:1996ff}. }
\label{tab:pi2Decay}
\end{table}

\subsection{Summary}

We have presented a computation of light hybrid decays and structure that assumes a simple constituent gluon can capture much of low lying gluodynamics. This approach finds support in lattice field computations, which determine that an axial gluonic degree of freedom is sufficient to describe the main features  of the low lying hybrid spectrum. The results are in  agreement with lattice decays obtained at the SU(3) flavor point for the $J^{PC}=1^{-+}$ hybrid, lending support to the model. We find, however, some model-dependence in extrapolating to physical masses for the $1^{-+}$ hybrid, whereas the other hybrids in the $H_1$ multiplet are less sensitive to this extrapolation. 

\begin{figure}[h]
    \centering
    \includegraphics[width=0.8\textwidth]{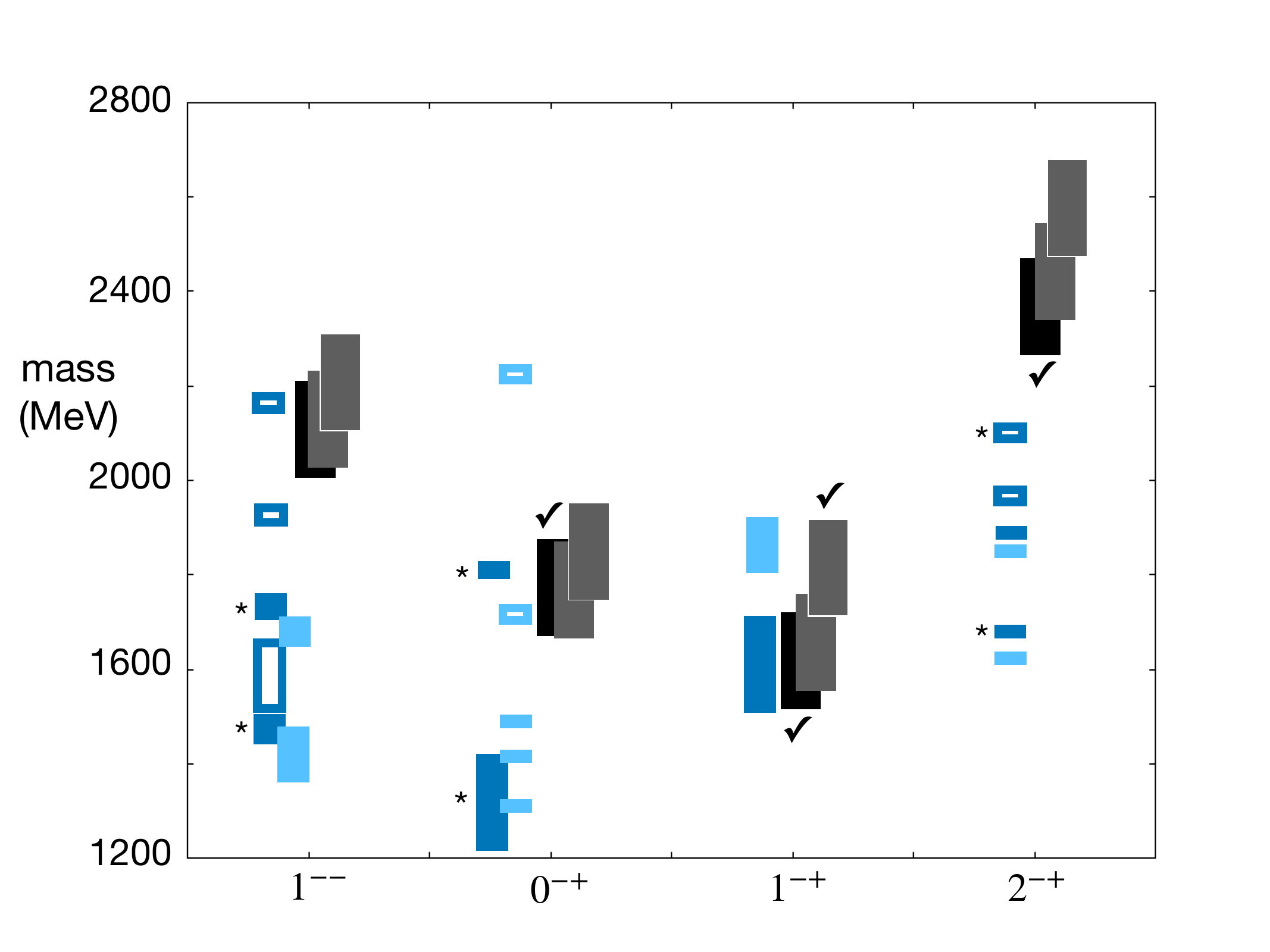}
    \caption{Estimated $H_1$ Hybrid Spectrum and Experimental Masses. Black boxes: lattice estimate isovector hybrid, gray boxes: lattice estimate isoscalar hybrid, blue boxes: PDG isovector, light blue box: PDG isoscalar. Box heights indicate errors in mass estimates, hollow boxes denote unverified states. Stars indicate isovectors that correspond well to quark model predictions. Suggested observed states are indicated with check marks.}
    \label{fig:H1}
\end{figure}

The ``extrapolated" lattice hybrid spectrum is compared to the experimental situation in Fig. \ref{fig:H1}. Stars in the figure indicate states that roughly correspond to quark model expectations -- showing that the agreement between model and experiment can be tenuous higher in the spectrum. Nevertheless, the points of contact between experiment and the hybrid spectrum, provided by $\pi_1(1600)$, $\eta_1(1855)$, and possibly $\pi(1800)$, point to the possible  emergence of a light hybrid spectrum.

The results presented here imply that the isovector vector hybrid is very broad, and therefore unlikely to be directly visible in production experiments such as $e^+e^- \to $ hadrons. Alternatively, the $\pi(1800)$ is a good candidate for the $H_1(0^{-+})$ hybrid with a distinctive pattern of decay modes. If verified, this implies that a narrow $s\bar{s}g$ state, $\eta(1900)$, should exist.

At present, the $\pi_1(1600)$ signal is the leading candidate for a hybrid meson. The consistency of its mass with the $\pi(1800)$ and the newly observed $\eta_1(1855)$, which can be interpreted as its $s\bar{s}g$ isoscalar partner state, adds confidence to claims that a resonance is present with exotic quantum numbers. An implication of this study is that a light-quark $\eta_1$ isoscalar should be sought with mass near 1750 MeV and a total width of approximately 150 MeV.

We have suggested that the $H_1(\pi)$ hybrid meson is already known, but remains to  be disentangled from the $\pi(3S)$ state.

 Finally, there is tantalyzing evidence for the
  isovector $H_1(\pi_2)$ in COMPASS $p \pi \to 3\pi p$ data. We  strongly suggest determining if this is indeed as resonance, perhaps by measuring its phase with respect to the $J^{PC}=0^{-+}$ channel, and then 
  examining quantities such as the $f_2\pi/\rho\pi$ partial width ratio or S to D amplitude ratios in the $f_2\pi$ decay mode  to establish its exotic character. In general, we expect that the isovector and isoscalar  $J^{PC}=2^{-+}$ hybrids are  narrow, with total widths between 60 and 80 MeV.

\acknowledgments
The authors thank Matt Shepherd for bringing Ref. \cite{COMPASS:2015gxz} to their attention.
Swanson acknowledges support by the U.S. Department of Energy under contract DE-SC0019232. This work contributes to the goals of the US DOE ExoHad Topical Collaboration, Contract No. DE-SC0023598.

\end{document}